\documentclass[showpacs,twocolumn,amsmath,amssymb,floatfix,pre,aps,nofootinbib]{revtex4}
\usepackage{graphicx}
\usepackage{amssymb}

\def\sn{\textrm{sn}}
\def\cn{\textrm{cn}}
\def\dn{\textrm{dn}}
\def\cs{\textrm{cs}}
\def\sc{\textrm{sc}}
\def\ds{\textrm{ds}}

\def\K{{K}}
\def\real{\textrm{Re}}

\begin{document}

\title{Triply-periodic Smectics}

\author{Christian D. Santangelo}
\altaffiliation{Future address: Department of Physics, University of Massachusetts, Amherst, MA 01003-9337, USA.}
\affiliation{Department of Physics and Astronomy, University of Pennsylvania, Philadelphia, PA 19104-6396, USA}
\author{Randall D. Kamien}
\email{kamien@physics.upenn.edu}
\affiliation{Department of Physics and Astronomy, University of Pennsylvania, Philadelphia, PA 19104-6396, USA}

\date{\today}

\begin{abstract}
Twist-grain-boundary phases in smectics are the geometrical analogs of the Abrikosov flux lattice in superconductors.  At large twist angles, the nonlinear elasticity is important in evaluating their energetics.  We analytically construct the height function of a $\pi/2$ twist-grain-boundary phase in smectic-A liquid crystals, known as Schnerk's first surface.  This construction, utilizing elliptic functions, allows us to compute the energy of the structure analytically.  By identifying a set of heretofore unknown defects along the pitch axis of the structure, we study the necessary topological structure of grain boundaries at other angles, concluding that there exist a set of privileged angles and that the $\pi/2$ and $\pi/3$ grain boundary structures are particularly simple.
\end{abstract}


\maketitle

\section{Introduction}

Because of their stability and quantization, it is natural to regard topological defects as independent degrees of freedom in systems with broken symmetry.  Perhaps the most storied example is the Abrikosov phase of type II superconductors \cite{abrikosov}.  There, in the London limit, variations in the phase of the macroscopic wave function can be decomposed into a ``smooth'' part and a singular part, the former being analogous to spin waves.  The singular component represents the topological defects -- those lines (in the three dimensional case) around which the phase slips by $2\pi$.  The spin waves can be removed from the theory resulting in an effective theory of repulsive vortices \cite{ChaikinnLubensky}.  Being the ideal proving ground for the study of broken symmetries, liquid crystalline states provide us with a variety of line-like and point-like defects which, because of the anisotropy of the surrounding medium, can enjoy orientationally dependent interactions.  With the flux-lattice in mind, we consider the smectic-A liquid crystal which is a close analog of the superconductor \cite{deGennes}.  Indeed, both the superconductor and the smectic-A phase have a complex scalar order parameter, representing macroscopic phase ordering of the Cooper pair wavefunctions and the one-dimensional periodic density modulation, respectively.  The superconducting order parameter is
minimally coupled to a gauge field, while the smectic-A order parameter is coupled to the director modes of a nematic phase which exists at higher temperatures:
\begin{equation}
F=F_{\rm smA}[\psi,{\bf n}] + F_{\rm Frank}[{\bf n}]
\end{equation}
where 
\begin{equation}\label{eq:sma}
F_{\rm smA}[\psi,{\bf n}] = \int d^3\!x\,\left\{\left\vert\left(\nabla - i q_0{\bf n}\right)\psi\right\vert^2 + r \vert\psi\vert^2 + u\vert\psi\vert^4\right\} \end{equation}
and $F_{\rm Frank}$ is the standard Frank free energy for a (possibly chiral) nematic:
\begin{eqnarray}\label{eq:frank}
F_{\rm Frank}[{\bf n}]&=& \int d^3\!x\,\Big\{K_1 \left(\nabla \cdot {\bf n}\right)^2 + K_2 \left[ {\bf n} \cdot \left(\nabla \times {\bf n} \right) + k_0 \right]^2\nonumber\\
&&+ K_3 \left[{\bf n} \times \left(\nabla \times {\bf n} \right) \right]^2 \Big\}
\end{eqnarray}
and where ${\bf n}$ is the unit director field.
Though deceptively similar to, for instance, a gauge-fixed Landau-Ginzburg theory the differences are profound.  Fluctuation effects are famously more complicated  at both the transition \cite{LubNem} and in the ordered state \cite{GrinPel,TRM,Milner}.

The Landau-Ginzburg-De Gennes theory predicts a nematic to smectic-A transition when $r<0$.  In the ordered state $\vert\langle\psi\rangle\vert\ne 0$ and the gradient term in (\ref{eq:sma}) favors states for which $\nabla\psi = iq_0{\bf n}\psi$. Because ${\bf n}$ is a unit vector, the phase of $\psi$ necessarily varies in space.  In the ground state, if we take the nematic to order along $\hat z$, then $\psi=\vert\psi\vert e^{iq_0 z}$.  In the ``London'' limit, we consider fluctuations only in the phase
of $\psi$ and not $\vert\psi\vert$.  Writing $\psi=\vert\psi\vert \exp\{iq_0[z-u(x,y,z)]\}$, $\hat{n} = \hat{z} - \delta \vec{n}$, and expanding $F_{\rm smA}$ to quadratic order in the fluctuations $u$ and $\delta \vec{n}$, we have:
\begin{eqnarray}\label{eq:quad}
F&\approx&{1\over 2}\int d^3\!x\, \Bigg\{B\left(\partial_z u\right)^2 + B\left(\nabla_\perp u - \delta {\vec n}\right)^2\nonumber\\ &&~\qquad\qquad+ K_1\left(\nabla_\perp\cdot\delta{\vec n}\right)^2 + K_2\left(\nabla_\perp\!\times\!\delta{\vec n}\right)^2 \nonumber\\ &&~\qquad\qquad+ K_3\left(\partial_z \delta{\vec n}\right)^2+2K_2k_0\nabla_\perp\!\times\!\delta{\vec n}\Bigg\}
\end{eqnarray}
where $\delta \vec{n}$ is the projection of the director in the xy-plane (appropriate for small director fluctuations), $B=2q_0^2\vert\psi\vert^2$, and $k_0$ is a pseudoscalar which would set the cholesteric pitch.  Even in this quadratic theory, we can see that the transverse mode of $\delta\vec n$ decouples from the (Eulerian) layer displacement $u(x,y,z)$.  The gauge-like coupling of $\nabla_\perp u$ to $\delta\vec n$ sets $\delta\vec n=\nabla_\perp u$ at long distances.  As a result, the transverse modes of the director are attenuated at a length scale $\lambda=\sqrt{K/B}$ where $K$ is on the order of the Frank elastic constants.  Known as the Meissner effect in superconductors this result shows that smectic order excludes twist, $\nabla_\perp\!\times\!\delta\vec n$.

Departing from the quadratic approximations of (\ref{eq:quad}) we can make this observation more general.  In the full nonlinear theory, $\langle\,\vert\psi\vert\,\rangle$ is still nonvanishing when $r<0$.  In the London limit we write $\psi=\vert\psi\vert e^{iq_0\Phi}$ where $\Phi(x,y,z)$ is a phase field and the mass density 
is   $\rho(\textbf{x}) = \rho_0 + \vert\psi\vert \cos [q_0 \Phi(\textbf{x})]$.    It follows that the smectic layers sit at the density peaks defined by $q_0\Phi({\bf x})=2\pi n$ for $n\in{\mathbb{Z}}$, {\sl i.e.} level sets of $\Phi({\bf x})$.  Again, the gradient term in (\ref{eq:sma}) requires that $\nabla\Phi = {\bf n}$ for constant $\vert\psi\vert$, so at long distances the director is perpendicular to the level sets or, in other words is parallel to the unit layer normal ${\bf N}=\nabla\Phi/\vert\nabla\Phi\vert$.   However, the twist order parameter is $k={\bf n}\cdot\left(\nabla\!\times\!{\bf n}\right)$, and it is straightforward to check that if ${\bf n}\propto\nabla\Phi$ then $k=0$. Thus, the smectic layers are incompatible with twist\footnote{The converse holds as well.  Namely, if ${\bf n}\cdot\left(\nabla\!\times\!{\bf n}\right)=0$ then surfaces can be found with $\bf n$ as their unit normals \protect\cite{Flanders}}.  

In order to relieve the incompatibility between smectic order and twist, it is necessary to allow $\vert\psi\vert$ to vary and, in fact, vanish at isolated points to form topological defects.  Indeed, the competition between molecular chirality and the existence of layers leads to the celebrated twist-grain-boundary (TGB) phase~\cite{renn, zas}.  This phase is the analog of the Abrikosov flux lattice~\cite{abrikosov}, with screw dislocations replacing flux lines and molecular chirality replacing the magnetic field.  In smectics-A, however, there is an additional complication: the coupling of geometry to elasticity requires that the screw dislocations and layers rotate together.
For small angles of rotation, the layer structure can be approximated using linear elasticity (\ref{eq:quad}) \cite{renn, bluestein2}.  The underlying rotational invariance of the compression strain, however, necessitates certain essential nonlinearities which have profound effects on the ground state energetics and layer  displacements \cite{bluestein, kamien, bps}.  For large-angle twist-grain boundaries, seen for instance in bent-core systems \cite{clark}, the nonlinearities become important and we are forced to confront the full nonlinear elasticity.   Because of the difficulty in systematically removing the unit director $\bf n$ from the theory, we go immediately to a rotationally-invariant free energy
in terms of the phase field $\Phi({\bf x})$.  Again, the gradient coupling in (\ref{eq:sma}) sets $\nabla\Phi = {\bf n}$, so we know that the compression strain vanishes when $\vert\nabla\Phi\vert=1$ or, equivalently, when ${\bf N}\cdot\nabla\Phi=1$, which indicates that the layers are spaced by one period along the layer normal -- {\sl i.e.} equally spaced layers.  We write 
the compression strain in terms of both $\Phi$ and the Eulerian displacement field $u({\bf x})=z -\Phi({\bf x})$:
\begin{equation}
\label{eq:strain}
u_{z z} = \frac{1}{2} \left[1 - \left( \nabla \Phi\right)^2\right] = \partial_z u - \frac{1}{2} \left(\nabla u\right)^2.
\end{equation}
The factor of $1\over 2$ is introduced so that in the linearized strain, $u_{zz}=\partial_z u$, the standard result.  Note that the nonlinear term in
(\ref{eq:strain}) is required by rotational invariance and is not merely an anharmonic correction to the elasticity.  It is responsible for strong and subtle corrections to linear elasticity~\cite{bluestein, kamien, bps}.  The bending energy is inherited from the Frank free energy and is simply $\nabla\cdot{\bf N}$.  We note, however, that this is precisely twice the mean curvature of the layers~\cite{rmp}, $H={1\over 2}\nabla\cdot{\bf N}$.  Finally, we write the full nonlinear free energy as 
\begin{equation}
\label{eq:free}
F = \frac{B}{ 4}\int d^3\!x\left\{ \left[\left(\nabla\Phi\right)^2-1\right]^2 +  8\lambda^2 H^2\right\}
\end{equation}
where $B$ is the compression modulus and $\lambda^2\equiv K_1/B$ is the ``splay penetration length''.  
The challenge is to find ground states which minimize (\ref{eq:free}) with the appropriate boundary conditions.  At infinity we take the layers to be perpendicular to the $z$-axis so that $\lim_{{\bf x}\rightarrow\infty}\nabla\Phi =\hat z$.  There are other boundaries, however, namely lines corresponding
to topological defects where $\Phi$ changes by $2\pi/q_0$.  

\begin{figure}[t]
\begin{center}
\resizebox{3in}{!}{\includegraphics{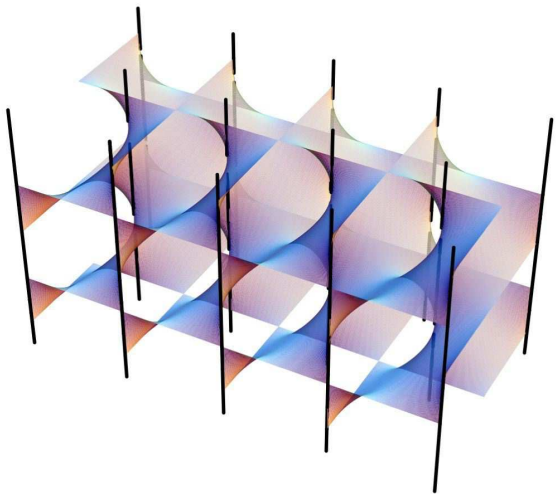}}
\caption{Schnerk's first surface, with charge $+2$ and $-2$ screw dislocations.  Note that the $-2$ dislocations lie at the center of the rectangle made by the adjacent $+2$ dislocations.   We choose $\theta=\psi$ and $k^2\approx  -0.03033$ so that $\K'(k)/\K(k) = 2-i$.}
\label{fig:schnerkpic}
\end{center}
\end{figure}

Recently, we have developed an approach to study TGB phases with $\pi/2$ grain boundaries~\cite{santangeloPRL}.  Our construction relies on a duality: $\pi/2$ twist-grain boundaries can be equivalently constructed from a sum of dislocations with burgers vector $b$, or a sum of dislocations rotated by ninety degrees with burgers vector $-b$.   Using this duality, a closed-form expression for the height of the smectic layers ensues and generates Schnerk's first surface (see Fig.~\ref{fig:schnerkpic}), one of a class of mathematical surfaces formed by summing individual screw dislocations.  In this paper, we elaborate on our construction and the energetics of Schnerk's first surface.  We then identify a third set of defects that lie along the pitch axis.  We use this set of previously unknown defects to study rotation angles smaller than $\pi/2$ and identify a privileged set of angles with particularly simple structures and correspondingly low energies.

The building blocks of any TGB phase are the screw dislocations -- when properly put together these effect a twist on the smectic layers without requiring the complete disappearance of the smectic order along a two-dimensional wall.
We write the screw dislocation in terms of the phase field $\Phi_h$ as:
\begin{equation}\label{eq:helicoid}
\Phi_h(\textbf{x}) = z - \frac{b}{2 \pi} \tan^{-1} \left(\frac{y}{x} \right).
\end{equation}
It is easy to verify that this phase field is an extremum of equation (\ref{eq:free}).  In addition, for $b = \pm 2$, this surface is identical to a relic minimal surface ($H=0$)~\cite{nitsche}, the helicoid, and is harmonic in both three dimensions ($\nabla^2 \Phi_h = 0$) and in the $xy$-plane ($\nabla_\perp^2\Phi=0$).  
Minimal surfaces have long been used as {\sl ansatzen} for smectic structures because they are global minima of the bending energy~\cite{hoffman}.  Nonetheless, the interplay between the compression and bending energies ensures that true minimizers of equation (\ref{eq:free}), with more complicated topologies than the helicoid, will neither have a harmonic phase field nor be minimal.

In this paper, we will take the opposite tack and consider, not minimal surfaces, but harmonic $\Phi$ formed by taking arbitrary sums of parallel screw dislocations.  If we take the dislocations to have their defect axes along the $\hat{z}$ direction with positions $\textbf{x}_i = (x_i, y_i)$ in the $xy-$plane, then we write
\begin{equation}\label{eq:phi}
\Phi(\textbf{x}) = \gamma z - \sum_i \Phi_h(\textbf{x}-\textbf{x}_i).
\end{equation}
As we will demonstrate, these sums will allow us to generate models for $\pi/2$ TGB phases.  Furthermore, there is an unexplored but close connection between surfaces generated by equation (\ref{eq:phi}) and certain triply- and doubly-periodic minimal surfaces.

This paper is organized as follows.  In section~\ref{sec:singleTGBs}, we review the nonlinear theory of large-angle twist-grain boundaries, and establish our notation.  In section~\ref{sec:surface}, we present our theory for the $\pi/2$ TGB structure based on Schnerk's first surface.  In section~\ref{sec:properties}, we discuss the properties of Schnerk's first surface.  In section~\ref{sec:moire}, we discuss the possibility of a moir\'e phase~\cite{moire} of screw dislocations in TGB structures of arbitrary angle and discuss a potential lock-in mechanism for certain large angles.  Finally, in section~\ref{sec:discussion} we summarize our results.

\section{Twist-grain boundaries and their duals}
\label{sec:singleTGBs}

Here we review the construction of a single twist-grain boundary with arbitrarily large angles.  To simplify our analysis, we assume the smectic layers have normal ${\bf N}= \hat{z}$ in the absence of dislocations and introduce the complex variable $w = x + i y$ for the coordinates along the plane normal to ${\bf N}$.  In this notation, a single screw dislocation of Burgers scalar $b$ is given by
\begin{equation}
\Phi_h(\textbf{x}) = \gamma z - \frac{b}{2 \pi} \textrm{Im} \ln (w).
\end{equation}
We introduce the constant $\gamma$ here, which is set by enforcing the boundary condition that the compression strain vanish at infinity.  In this simple case of one screw dislocation, we set $| \nabla \Phi | \rightarrow 1$ at infinity, implying that
\begin{equation}
\gamma^2 = 1 - \lim_{w \rightarrow \infty} \left[ \frac{b}{2 \pi |w|} \right]^2 = 1.
\end{equation}
As we shall see, when considering doubly- and triply-periodic smectics, this boundary condition will depend on the orientation and position of the screw dislocations.

A single twist-grain boundary can be decomposed into a set of dislocations with spacing $\ell_d$ and positions $w + \ell_d n$ for integers $n$~\cite{renn, kamien}.  Though we know of no minimizer of equation (\ref{eq:free}) with this property, progress can be made by considering the sum~\cite{kamien}
\begin{equation}
\label{eq:TGBsum}
\Phi_{\rm row}(z,w) = \gamma z - \frac{b}{2 \pi} \sum_{n=-\infty}^\infty \textrm{Im} \ln \left(w + \ell_d n \right).
\end{equation}
Utilizing the infinite product $\sin w = w\prod_{n\ne 0} \left(1-\frac{w}{ n\pi}\right)$ it follows that \cite{kamien2}
\begin{equation}\label{eq:TGBsine}
\Phi_{\rm row} = \gamma z-\frac{b}{ 2\pi}\textrm{Im}\ln\sin\left(\frac{\pi w}{ \ell_d}\right).
\end{equation}
To set $\gamma$ now, we must go out along the $y$-axis only, since the $x$-axis contains
the grain boundary.  
Setting $|\nabla \Phi_{\rm row}| \rightarrow 1$ as $y \rightarrow \pm \infty$, we have
\begin{equation}
\gamma^2 = 1 - \left(\frac{b}{2 \ell_d}\right)^2.
\end{equation}
The layers are flat at  $y = -\infty$ with layer normal given by $\textbf{N}_- = - \frac{b}{2 \ell_d} \hat{x} + \gamma \hat{z}$.  For $y = +\infty$, the layer normal is $\textbf{N}_+ = \frac{b}{2 \ell_d} \hat{x} + \gamma \hat{z}$.  This gives an overall angle of rotation $\alpha=2\sin^{-1}\left[ b/(2\ell_d)\right]$~\cite{renn,kamien}.

This structure has the same topology as Scherk's first surface \cite{nitsche}, a minimal surface which has been used as a conjectured structure for TGBs in diblock copolymers~\cite{hoffman}.  Surprisingly, this sum and Scherk's first surface are merely uniform dilations of each other: the level sets of $\Phi_{\rm row}(x,y \cos(\alpha/2),z)$ are {\sl identical} to Scherk's first surface~\cite{kamien2} when $b=2$.  Furthermore, the rescaled surface is minimal for any $b$ despite the fact that the mean curvature is nonlinear.  It is reasonable to conjecture that the true smectic layer geometry falls somewhere in between the exact sum of dislocations and Scherk's surface, and this expectation has been borne out by recent numerical studies of both lyotropic~\cite{schick} and thermotropic smectics~\cite{TGBsim}.
Numerically, $\Phi_{\rm row}$ (without rescaling) is a slightly better than Scherk's surface as a model for the layers.  An important caveat, however, is that at large angles the director decouples from the layer normal near the dislocation cores~\cite{TGBsim}, and it would be necessary to reintroduce director modes to make contact with simulation.  Nevertheless, approximate analytical models are useful guides for our understanding and, in this case, yield insights into the TGB smectic structure.

The coordinates $(x,y,z)$ of the level set $\Phi_{\rm row}=0$ satisfy \cite{kamien}
\begin{equation}\label{eq:levelsetTGB1}
\tan(2\pi\gamma z/b)\tan(\pi  x/\ell_d)=\tanh(\pi y/\ell_d).
\end{equation}
Equation (\ref{eq:levelsetTGB1}) possesses a hidden symmetry: if we rotate the entire structure by $\pi/2$, interchange $(x,z) \rightarrow (z,-x)$, and simultaneously take $b \rightarrow -b$, the equation is invariant.  For a general rotation angle $\alpha$, this transformation exchanges $\ell_d$ and $b/(2 \gamma)$ in equation (\ref{eq:levelsetTGB1}) and changes the rotation angle from $\alpha$ to $\pi - \alpha$.  This allows us to view the level sets of a single TGB of any angle $\alpha$ as being constructed either from parallel defects with burgers scalar $b$ (along $\hat z$) or from parallel defects with burgers scalar $-b$ rotated by $\pi/2$ (along $\hat x$).

On the level sets of $\Phi_{\rm row}$, we calculate
\begin{eqnarray}
\nabla \Phi_{\rm row} & &= \frac{2 \pi \gamma}{b} \frac{\cos^2(2 \pi \gamma z/b)}{\tan(\pi x/\ell_d)} \Bigg\{ \frac{2 \pi \gamma \tan(\pi x/\ell_d) \hat{z}}{b \cos^2(2 \pi \gamma z/b)}\nonumber\\
& &+ \frac{\pi \tan(2 \pi \gamma z/b) \hat{x}}{\ell_d \cos^2(\pi x/\ell_d)}+ \frac{\pi \hat{y}}{\ell_b \cosh^2(\pi y/\ell_b)} \Bigg\}.\label{eq:gradphiTGB}
\end{eqnarray}
For $\alpha = \pi/2$, when $2 \pi \gamma/b = \pi/\ell_d$, it is clear from equation (\ref{eq:gradphiTGB}) that $(x,z) \rightarrow (z,-x)$ preserves the unit normal vector ${\bf N}$.
The magnitude of $\nabla \Phi_{\rm row}$, however, is not preserved by this transformation.  To see why this must be the case, notice that $| \nabla \Phi_{\rm row} |$ diverges at the core of defects along the $\hat{z}$ axis, even though we have now identified an additional set of helicoids along the $\hat{x}$ axis.  Though rotations by ninety degrees preserves the \textit{level sets} of the surfaces, as indeed they must, they need not preserve the structure of the phase field between those level sets and, in particular, do not preserve the orientation of the defect cores.  We conclude that only one set of helicoids need have cores.  However, the total compression energy and bending energy remains unchanged by a ninety degree rotation -- the direction of the defect cores does not alter the energy.

\section{Schnerk's first surface}
\label{sec:surface}
\subsection{The $\pi/2$ TGB phase structure}\label{sec:piover2}
A $\pi/2$ TGB phase consists of twist-grain boundaries with defects alternating along the $\hat z$ and $\hat x$ directions (we assume $\hat y$ is the pitch axis) and with rotation angle $\alpha = \pi/2$.  Finding an analytical expression for such a structure from which the energy can be calculated is a daunting task~\cite{bluestein}.  However, the degeneracy in how we identify the defects in a single TGB structure allows a significant simplification in the structure of this phase.  By applying the appropriate ninety degree rotations, all of the dislocations can be rotated to be parallel to the $z$ axis, leading to a structure of alternating TGBs such as the one shown in Fig.~\ref{fig:schnerkpic}.  This allows us to compute analytically the level sets of a $\pi/2$ TGB phase by summing parallel, alternating TGBs separated by a distance $\ell_b$.  This sum is given formally by
\begin{equation}\label{eq:schnerksum}
\Phi_{\rm TGB}(\textbf{x}) = \gamma z - \frac{b}{2 \pi} \textrm{Im} \sum_{m=\infty}^\infty (-1)^m \ln \sin \left(\frac{\pi w}{\ell_d} + m \frac{\pi \tau}{2} \right),
\end{equation}
where $\tau$ is a complex number that generates the appropriate translation between grain boundaries.  To ensure that each grain boundary rotates the layers by $\pi/2$, we might set, using the results of the previous section, $b/(2\ell_d) = \sin(\pi/4)=1/\sqrt{2}$.  However, this relation requires a modification owing to the collaborative effect of the adjacent grain boundaries.  We put off the details of 
setting the rotation angle to later.  For now, we keep $\ell_d$ and $b$ as free parameters.   

We may rewrite the infinite sum in $\Phi_{\rm TGB}$ 
(\ref{eq:schnerksum}) 
as 
\begin{equation}
\Phi_{\rm TGB}({\bf x})=\gamma z - \frac{b}{2 \pi} \hbox{Im}\,\ln\Theta(w)
\end{equation}
where
\begin{eqnarray}
\Theta(w)=\prod_{m~\textrm{even}} \frac{\sin \left(\pi w/\ell_d + m \pi \tau/2 \right)}{\sin \left[ \pi w/\ell_d + (m+1) \pi \tau/2 \right]}\nonumber
\end{eqnarray}
Because $\Phi_{\rm TGB}$ is a sum of harmonic functions in $x$ and $y$, it follows that $\ln\Theta(w)$ is
analytic as is $f(w)=e^{\Theta(w)}$.  Moreover, $f(w)$ is doubly periodic and it thus follows that
$f(w)$ can be represented in terms of elliptic functions.  Indeed, 
through an appropriate rescaling of $x$ and $y$, $f(w)$ shares all the poles and zeroes of the Jacobi elliptic function $\sn(u,k)$~\cite{elliptic}.  The same result can be established through one of the infinite product formulas for $\sn(u,k)$.  We arrive at the exact summation of screw dislocations for
a $\pi/2$ TGB structure:
\begin{equation}\label{eq:infiniteproduct}
\Phi_{\rm TGB}(\textbf{x})=\gamma z - \frac{b}{2 \pi}\hbox{Im}\ln \sn \left[\theta x + i \psi y,k\right]
\end{equation}
where $\theta$ and $\psi$  are the necessary scale factors, $\K(k) = \int_0^1dx [(1-x^2) (1- k^2 x^2)]^{-1/2}$ is an elliptic period, $i\K'(k) =\int_1^{1/k}dx [(1-x^2) (1- k^2 x^2)]^{-1/2}= i\K(\sqrt{1-k^2})$ is the other elliptic period, and $k$ is the elliptic modulus.

The ratio of the elliptic periods is $\tau\equiv i\K'(k)/\K(k)$.  It is particularly simple to consider the case that $k$ is pure imaginary so that $k^2<0$.  In this case
$\textrm{Im}\,{\K'(k)} = - \K(k)$ (see the Appendix), $\tau= \hbox{Re}\;\K'(k)/\K(k) +1$ and so we set $\theta \equiv 2 \K(k)/\ell_d$,  $\psi \equiv \real \K'(k)/\ell_b$ to achieve the desired periodicity.  
Though we could tune the real part of $\tau$, which controls the offset between grain boundaries, to generate a family of surfaces with different topologies,
these identities
allow for a particularly straightforward analysis of the energetics of Schnerk's first surface.  Other choices of $\tau$ lead to elliptic moduli in which $k$ is, in general, complex.

The level sets of $\Phi_{\rm TGB}$, Schnerk's first surface, generate the triply-periodic surface shown in Fig.~\ref{fig:schnerkpic} for $\ell_b = \ell_d$.  Though Schnerk's first surface is not a minimizer of the smectic free energy, equation (\ref{eq:free}), it's construction ensures that $\Phi_{\rm TGB}$ is at least a harmonic function.  In analogy with a single grain boundary which is topologically Scherk's first surface, Schnerk's first surface is likewise topologically identical to the Schwarz D surface, another minimal surface.  This can be seen by comparing the unit cell of Schnerk's first surface (see Fig. 2 of ref.~\cite{santangeloPRL}) with that of the Schwarz D surface.  Unfortunately, we are not aware of a simple transformation, such as a rescaling of one or more of the coordinates, that renders Schnerk's surface exactly minimal.  It may be interesting to note that the Schwarz D surface also has a parametric representation in terms of elliptic functions~\cite{weellip}.

\subsection{Schnerk's first surface and the Jacobi elliptic functions}
For arbitrary sums of parallel screw dislocations, $\eta(w)=\partial_x \Phi - i \partial_y \Phi$ is a meromorphic function of $w$ whose simple poles correspond to screw dislocations.  In the case of Schnerk's first surface, $\eta$ is a doubly periodic function of $w$ and we are immediately led to consider generalizations of Schnerk's first surface to other defect lattices.  Fortunately,  by Liouville's theorem, a doubly periodic, meromorphic function must have residues which add to zero\footnote{Consider a closed contour around the boundary of one period in the complex plane.  By periodicity the closed contour integral vanishes and thus the sum of the 
residues must vanish.}, which in our case means net charge neutrality of the screw dislocations.  Thus our construction, or any generalization of it, must always generate {\sl achiral} phases since the net twist always vanishes.  It is only because of the special duality at $\alpha=\pi/2$ that we can construct a rotating structure (though the rotation can be thought of in either direction).    For alternating TGBs at other twist angles, adjacent boundaries rotate the layers in opposite directions.
This observation and Liouville's theorem provides an alternative derivation of an observation by Sethna, found in reference~\cite{renn}, that constant density configurations of parallel screw dislocations with the same charge are impossible.  In fact, a sum of screw dislocations with the \textit{same} charge can be performed formally, after a suitable regularization, in terms of the Weierstra\ss\ elliptic function, $\zeta(w)$~\cite{elliptic}.  This function, however, fails to be doubly-periodic and $\eta(w)$ fails to describe a suitable layer geometry.  We will return to the question of describing TGB phases with other twist angles in the next section.

We also point out that limiting the discussion to $k^2<0$ is not too restrictive a condition.  By utilizing the vast array of elliptic function identities, one can rewrite many choices of $\tau$ (and subsequently $k$) in terms of an algebraic combination of elliptic functions with $k^2<0$.
Notice, for example, that~\cite{ellipticreference}
\begin{equation}
\sn(w,i k) = \frac{\sn(w \sqrt{1+k^2}, k/\sqrt{1+k^2})}{\dn(w \sqrt{1+k^2}, k/\sqrt{1+k^2})},
\end{equation}
allowing us to relate elliptic functions with $k^2<0$ to those with $k^2>0$.  Similar relations map $\cn$ to $\cn/\dn$ and $\dn$ to $1/\dn$.  Other relations allow us to map elliptic functions with modulus $k$ to those with modulus $1/k$, and the Landen transforms yield more complicated identities between elliptic functions with different moduli.

\section{Properties of Schnerk's first surface}
\label{sec:properties}
\subsection{Energetics}
To simplify our notation, we define $\zeta = \theta x  + i\psi y$ (not to be confused with the Weierstra\ss\ elliptic function), denote its complex conjugate as $\bar\zeta$,  and use Glaisher's notation ($\cs$ for $\cn/\sn$, $\ds$ for $\dn/\sn$, {\sl etc.}) for the elliptic functions \cite{elliptic}, suppressing the elliptic modulus $k$.   Additionally, $\hbox{pq}(\bar\zeta) = \overline{\hbox{pq}(\zeta)}$ since $k^2$ is real.  The compression strain for $\Phi_{\rm TGB}$, $u_{zz}\equiv [1-(\nabla \Phi)^2]/2$, is
\begin{eqnarray}
u_{zz}&=& \frac{1}{2} \Big[1 - \gamma^2  - \frac{b^2}{8 \pi^2} \left(\theta^2 + \psi^2\right) \left\vert\cs\,\zeta \dn\,\zeta\right\vert^2\nonumber\\
& & +\frac{b^2}{8 \pi^2} \left(\theta^2 - \psi^2\right) \hbox{Re}\left[ \cs^2\zeta\; \dn^2\zeta \right]\Big]
\end{eqnarray}
Because we can choose the periodicity of our structure by either altering $k$ (and consequently $\tau$)or by  altering $\psi$ and $\theta$ we have some freedom when doing our calculations.  However, we note that though the symmetries are identical as we alter $k$ or $(\theta,\psi)$, the surfaces are not.  We will discuss this in the following.
For now, we note that if $\psi=\theta$, the compression strain is particularly simple.  For this choice, $k$ is determined through $i(1-\tau)=\textrm{Re} \K'(k)/\K(k) = 2 \ell_b/\ell_d$.

In the case of a single grain boundary, we set $\gamma$ by considering $y=\pm\infty$.  Here, the structure is triply periodic and we are forced to set $\vert\nabla\Phi\vert=1$  inside one of the periods to set $\gamma$.  We choose to have the compression vanish halfway between the grain boundaries, {\sl e.g.} along $y=\ell_b/2$ or $x=\ell_d/4$.  These lines are also where we choose to measure the rotation of the layers, and so this is a natural choice.  Though we should choose $\gamma$ to minimize the compression energy for a single periodic domain, as $\ell_b/\ell_d\rightarrow\infty$ these two procedures agree.   
Our choice of $k^2 < 0$ is, again, particularly useful in determining $\gamma$.  On the lines $\zeta = \K(k)/2 + it$ and $\zeta= t +i\real \K'(k)/2$, for $t\in\mathbb{R}$, it can be shown that $\vert \cs(\zeta,k)\,\dn(\zeta,k) \vert^2 = 1-k^2$ and thus $\vert\nabla\Phi\vert$ is constant as well.   We verify these identities in the Appendix.  We can, as a result, set $\gamma$ along these lines and find:
\begin{equation}
\gamma^2 = 1 -  b^2(1-k^2)\theta^2/(2\pi)^2
\end{equation}
Returning to the rotation of the layers, promised in Section \ref{sec:piover2}, we measure the angle
from $w=\ell_d/2 - i\ell_b/2$ to $w=\ell_d/2+ i\ell_b/2$.  We find:
\begin{equation}\label{twist}
\alpha = 2\sin^{-1} \left[b\sqrt{1-k^2}\theta/(2\pi)\right].
\end{equation}
As $k\rightarrow 0$, $\theta\rightarrow \pi/\ell_d$ and these expressions reduce to those for a single grain boundary.  Requiring $\alpha=\pi/2$ sets $\gamma^2=1/2$ and sets  $\ell_d=b\sqrt{2(1-k^2)}\K(k)/\pi$.  
Again, for $k^2>0$, or other generic complex values of $k$, the lines of constant tilt angle are no longer straight, making this procedure difficult, if not impossible.  

Finally, we make use of the expansion of Jacobi elliptic functions~\cite{jacobi} in terms of $q\equiv \exp\left[-\pi\K'/\K\right] = -\exp\left[-2\pi \ell_b/\ell_d\right]$:
\begin{widetext}
\begin{equation}\label{asym}
\ln\left[\sn\,\zeta\right] = \ln\sin\left(\frac{\pi w}{\ell_d}\right)+\sum_{m=1}^\infty \frac{2}{ m}\frac{q^m}{ 1+q^m} \cos\left(\frac{2m\pi  w}{\ell_d} \right)+\ln\left(\frac{2 q^{1/4}}{\sqrt{k}}\right)
\end{equation}
\end{widetext}
to compute the long distance interaction between grain boundaries.  Recall that in the case of a single grain boundary, the nonlinear strain (\ref{eq:strain}) leads to power-law interactions between defects \cite{kamien}.  Here, we are able to address the nature of the
interactions {\sl between} twist-grain boundaries.  Jacobi's formula (\ref{asym}) leads to a simple answer: the interactions are exponential.  To see this we note that the first term in (\ref{asym}) is the $w$-dependent part of the phase field for a single
grain boundary, as in (\ref{eq:TGBsine}).  Thus, were we to calculate the difference between the full TGB structure and a set of non-interacting grain boundaries, the correction would come from the infinite sum in (\ref{asym}) (note that constant $\ln(2q^{1/4}/\sqrt{k})$ drops out of the energetics).  Hence the corrections would be at least ${\cal O}(q)$ and $q=-\exp\left[-2\pi\ell_b/\ell_d\right]$: exponentially decaying interactions at $\ell_b$ grows with $\ell_d$ fixed.  In the symmetric case depicted in Fig. 1, though $\ell_b=\ell_d$, $q=-e^{-2\pi}\approx =-0.002$ is small enough for an expansion to be reliable.  We can compare this result to the linear elasticity theory.  There, without director modes, there is no interaction between screw dislocations.  With the director modes included, as in (\ref{eq:quad}), screw dislocations interact exponentially and it follows that grain boundaries will as well \cite{bluestein2}.  However, the attenuation length in this case is the twist penetration length $\lambda_T=\sqrt{K_2/B}$, and not the distance between defects, $\ell_d$.   Thus, the exponential interactions that we find here are different -- they arise from elastic strains and not from 
the optical modes of the director.   

There are some complications in calculating the interaction energy which are worth mentioning.  To calculate the interaction energy we use $\Phi_{\rm TGB}$ to evaluate the compression in one period, $\ell_b\times\ell_d$.  From this we subtract the energy
of $L_y/\ell_b$ isolated grain boundaries, where $L_y$ is the dimension of the system in the $y$ direction.  However, the isolated grain boundaries have tails which extend beyond a distance $\ell_b$.  Thus while the intensive interaction energy is the difference of two integrals, they are integrals over {\sl different} regions: the first $[-\ell_d/2,\ell_d/2]\times[-\ell_b/2,\ell_b/2]$, the second $[-\ell_d/2,\ell_d/2]\times[-\infty,\infty]$.   Fortunately, this does not spoil our argument that the interaction is a positive power of $q$.    In the tail from $\ell_b/2$ to $\infty$, the integrand arising from the single grain boundary falls off exponentially, also as $e^{-y/\ell_d}$ and
so the contribution from the tail is also proportional to $q$.  We arrive at:
\begin{equation}\label{interact}
\frac{\Delta F_c}{A}\sim \frac{BL_z\ell_d}{2\pi\ell_b}\left[C+\left(\frac{b}{2\pi\xi}\right)^2 +2\ln\left(\frac{2\sqrt{2}\xi}{b}\right)\right]q
\end{equation}
where $C$ is a positive constant of order unity whose precise value depends on the choice of cutoff at the dislocation core, $L_z$ is the $z$-dimension of the system, and an elastic cutoff length $\xi$ is introduced to cutoff a $\vert w\vert^{-4}$ divergence in $u^2_{zz}$ near the origin.   This cutoff was necessary in the case of a single grain boundary~\cite{kamien} as well and arises from the infinite periodicity at the core of a helix, {\sl i.e.} at the core of the defect the spacing between one sheet and the next vanishes.   Since $q=-\exp\{-2\pi\ell_b/\ell_d\}$, this is an attractive, exponential interaction.  Stepping back and examining the arrangement of defects, the attraction is not at all surprising.  The parallel defects in adjacent grain boundaries are of the opposite sign and would, left
to their own devices annihilate.  Because of this, we conjecture that even when $\theta\ne\psi$, the compression energy is attractive.  {\sl En passant}, we note that this form of the compression energy allows us to minimize over all possible values
of the core size.  Doing so, we find that $\xi\propto b$, reminiscent of Kleman's split core edge defects for large $b$ \cite{klemanbook,bps}.

We end this section by considering the bending energy.   For general $\theta$ and $\psi$, the mean curvature, $H=[(1+z_x^2) z_{y y} + (1 + z_y^2) z_{x x} - 2 z_x z_y z_{x y}]/(1 + z_x^2 + z_y^2)^{3/2}$ is the rather foreboding
\begin{widetext}
\begin{eqnarray}
H = \frac{b }{2 \pi \gamma} \Bigg\{ -\chi \hbox{Im}\left[ \frac{\cn^2\zeta \dn^2\zeta}{\sn^2\zeta}\right]  +2 \chi k^2 \hbox{Im}\,{\sn}^2{\zeta}  -
 2k^2 \frac{b^2}{4 \pi^2 \gamma^2} \theta^2 \psi^2\hbox{Im} \left( \frac{{\sn}^2\bar\zeta}{\sn^2\zeta}  \right) \Bigg\}/H_0^{3/2},
\end{eqnarray}
\end{widetext}
where
\begin{eqnarray}
H_0 &=& 1 + \frac{b^2}{2 \pi^2 \gamma^2} \left(\psi^2 - \theta^2\right) \hbox{Re}\, \cs^2\zeta \dn^2\zeta \nonumber\\
& & + \frac{b^2}{8 \pi^2 \gamma^2} \left(\psi^2 + \theta^2\right) \vert\cs\,\zeta \dn\,\zeta\vert^2,
\end{eqnarray}
and
\begin{equation}
\chi \equiv \theta^2 - \psi^2 - (1+k^2) {b^2 \over 4 \pi^2 \gamma^2} \theta^2 \psi^2.
\end{equation}
The mean curvature is finite everywhere and so, when calculating the energy of $\Phi_{\rm TGB}$, it is not necessary to introduce
a cutoff for the cores.  The energy may likewise be expanded in powers of $q$  and we again argue for an exponential interaction.  The choice $\psi=\theta$ achieved great simplicity in our analysis of $u_{zz}$ but the expression for $H$ does not suggest a simplifying choice.  However, the energy calculation does.  When subtracting the bending energy of the individual grain boundaries, a great simplification occurs if the surfaces are minimal, with $H=0$.  Recalling our discussion of the single grain boundary, if we set $\theta=\sqrt{2}\psi$ then the $k=0$ limit of  (\ref{eq:infiniteproduct}) is precisely Scherk's first surface, with $H=0$ everywhere \cite{kamien,kamien2}.  Thus, the interaction energy is positive definite and ${\cal O}(q^2)$ since $H$ is ${\cal O}(q)$. We are led to the appealing result that the purely repulsive bending energy can balance the attractive compression energy and set the preferred value of $\ell_b/\ell_d$.   

\begin{figure}[b]
\begin{center}
\resizebox{3in}{!}{\includegraphics{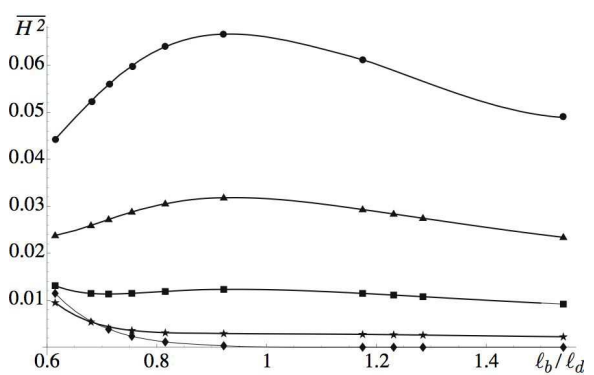}}
\caption{Average bending energy as a function of $\ell_b/\ell_d$. Results plotted are for $\psi = \theta/\sqrt{2}$ (diamonds), $\psi = \theta/1.3$ (stars), $\psi = \theta/1.2$ (squares), $\psi=\theta/1.1$ (triangles) and $\psi = \theta$ (circles); the lines are guides for the eye.  For $\psi=\theta/\sqrt{2}$, we explicitly recover the quadratic dependence on $q^2$ at small $q$.  In all cases, the interaction is repulsive for large $\ell_b/\ell_d\ge 1$ where the small $q$ limit holds.  All calculations are for $b=2$.}
\label{fig:bending}
\end{center}
\end{figure}

When $\psi = \theta$, a more natural choice from the point of view of the compression energy, the question of the sign of the bending energy interaction is much more difficult to answer without an involved calculation.  We have verified numerically, however, that the interaction remains repulsive with this choice of scale factors, as well as many others, at long distances.  In Fig. \ref{fig:bending}, we plot  the bending energy for several choices of ratios between $\psi$ and $\theta$, spanning from $\psi=\theta/\sqrt{2}$ to $\psi = \theta$.  Only for $\psi=\theta/\sqrt{2}$ do we find a purely repulsive bending energy for the entire range of $\ell_b$.  The cross-over from repulsive to attractive bending interaction occurs for $\ell_b/\ell_d \approx 1$, however, and we conclude that stable Shnerk phases can exist in cases that $\ell_b$ is large enough compared to the dislocation separation within a grain boundary.

\subsection{The triality of Schnerk's first surface}
We have already exploited the duality in the description of a single twist-grain boundary in constructing Schnerk's first surface.  Perhaps not surprisingly, Schnerk's first surface not only enjoys the original duality on which it was built but, in addition, an added symmetry.  Put together, Schnerk's first surface enjoys a somewhat Kafkaesque {\sl triality}~\cite{Kafka}.
The level sets of Schnerk's first surface satisfy
\begin{equation}\label{eq:parametricform}
\tan(2 \pi \gamma z/b) \frac{\sc[2 \K x/\ell_d]}{\dn[2 \K x/\ell_d]} = -i\frac{\sc[i\real\K' y/\ell_b]}{\dn[i\real\K'  y/\ell_b]}
\end{equation}
We recognize $\sc(\zeta)/\dn(\zeta)$ as the elliptic generalization of $\tan\zeta$, and note that it has a pole at $\zeta=\K$ of the same form as the pole in $\tan\zeta$ at $\frac{\pi}{2}$.  This allows us to view the surface as  being composed of defects along $x$, instead of $z$.  Near the zeroes in $y$ and $z$, let $\delta y=y-2n\ell_b$ and $\delta z =z -mb/(2\gamma)$ for $n,m\in\mathbb{Z}$ we have 
\begin{equation}
\frac{\sc[2\K x/\ell_d]}{\dn[2\K x/\ell_d]} \approx \frac{ \real\K' b}{2\pi\gamma\ell_b}\frac{\delta y}{\delta z}
\end{equation}
Likewise,  near the poles in $y$ and $z$, let $\delta y = y - (2n+1)\ell_b$ and $\delta z =z - (2m+1) d/4$, and
\begin{equation}\label{ugh2}
\frac{\sc[2\K x/\ell_d +\K]}{\dn[2\K x/\ell_d+\K]} \approx -\frac{\real\K' b}{2\pi\gamma\ell_b }\frac{\delta y}{ \delta z}
\end{equation}
Thus we see that we may view the Schnerk surface as made of oppositely charged defects staggered in the $yz$-plane.
In fact, this observation recapitulates the duality of Scherk's first surface we pointed out earlier.  Notice, however, that the rotation $(z,x) \rightarrow (x,-z)$ coupled with $b \rightarrow -b$ does not leave Schnerk's surface invariant (as it does for a single TGB).  

Further elliptic gymnastics demonstrates that the unmodified Schnerk's first surface may also be viewed as a charge neutral arrangement of defects in the $xz$-plane, with cores along the TGB pitch axis.  The poles are located at $x=(2 m+1) \ell_d/2$ and $z=(2n  + 1) b/(4\gamma)$ while the zeros at $x=m \ell_d$ and $z=n b/(2\gamma)$, for $m,n \in \mathbb{Z}$.   Near a zero in $x$ and a pole in $z$, we have  $\delta x = x - m\ell_d$ and $\delta z - z - (2n+1)b/(4\gamma)$, so that 
\begin{equation}
-i\frac{\sc[i\real\K' y/\ell_d]}{\dn[i\real\K' y/\ell_d]} \approx \frac{ -Kb}{\pi\gamma\ell_d}\frac{\delta x}{\delta z}
\end{equation}
Swapping the pole and the zero, we have
$\delta x=x -(2m+1)\ell_d/2$ and $\delta z = z - nb/(2\gamma)$:
\begin{equation}\label{ugh}
-i\frac{\sc[i\real\K' y/\ell_d]}{\dn[i\real\K' y/\ell_d]} \approx \frac{ \pi\gamma\ell_d}{\left(1-k^2\right) Kb}\frac{\delta z}{\delta x}
\end{equation}
where $k$ remains the elliptic modulus.  Fortuitously, $(1-k^2)\sc[u]/\dn[u] = \dn[u+i\real\K']/\sn[u+i\real\K']$ and so (\ref{ugh}) becomes
\begin{equation}
-i\frac{\sc[i\real\K' y/\ell_d+i\real\K']}{\dn[i\real\K' y/\ell_d+i\real\K']} \approx \frac{ Kb}{ \pi\gamma\ell_d}\frac{\delta x}{\delta z}
\end{equation}
we note that similar machinations were required in deriving (\ref{ugh2}), but we digress.
Again we see that we may view Schnerk's surface as being built of a lattice of alternating defects along the $y$, or pitch axis.  
Thus, we observe that Schnerk's first surface may be constructed from any one of \textit{three} orthogonal configurations of helicoid-like defects.  We will exploit these bonus dislocations along the pitch axis to find alternate topological constructions of arbitrary angle TGB phases. 

This discussion suggests utilizing a modified Schnerk surface given by the parametric construction~\cite{footnote1}
\begin{equation}\label{eq:generalSchnerk}
\textrm{sgn}(b) \frac{\sc[2 \K z/\ell_z]}{\dn[2 \K z/\ell_z]} \frac{\sc[2 \K x/\ell_d]}{\dn[2 \K x/\ell_d]} = i \frac{\sc[i\real\K' y/\ell_b]}{\dn[i\real\K'  y/\ell_b]}
\end{equation}
found by replacing $\tan(2 \pi z/d)$ with its elliptic generalization.  Here, we have replaced $|b|/\gamma$ with $\ell_z$ for notational simplicity.  When $\ell_z = \ell_d$, as is the case for Schnerk's first surface, it is straightforward to see that this surface exhibits the correct symmetry with respect to rotating the defects by $\pi/2$ while changing their sign.

Equation (\ref{eq:generalSchnerk}) exhibits an additional symmetry which we make manifest first by translating the surface along the $z$ direction by $\ell_z/2$.  This gives
\begin{equation}
\frac{\sc[2 \K x/\ell_d]}{\dn[2 \K x/\ell_d]} = i (1-k^2) \textrm{sgn}(b) \frac{\sc[2 \K z/\ell_z]}{\dn[2 \K z/\ell_z]} \frac{\sc[i \real\K' y/\ell_b]}{\dn[i \real\K' y/\ell_b]}.
\end{equation}
When $k^2=-1$ ($k=i$ and $2 \ell_b = \ell_d$), elliptic function identities yield the equivalent parametric form
\begin{equation}
\frac{\sc[2 \K x/\ell_d]}{\dn[2 \K x/\ell_d]} = -2 \frac{\sc[2 \K z/\ell_z]}{\dn[2 \K z/\ell_z]} \frac{\sc[\real\K' y/\ell_b]}{\dn[\real\K' y/\ell_b]}.
\end{equation}
From this it is clear that the translation $z \rightarrow z + \ell_z/2$ followed by the rotation $(z,y) \rightarrow (y,-z)$ in concert with $b \rightarrow -b$ leaves the surface given by equation (\ref{eq:generalSchnerk}) invariant for $\ell_z = \ell_b$.  
Finally, it is also possible to formulate this modified surface as a phase field
\begin{eqnarray}
\Phi(x,y,z) &=& \textrm{sgn} (b) \textrm{Im} \ln \left[\frac{\sc(2 \K z/\ell_z,k)}{\dn(2 \K z/\ell_z,k)} \right]\\
& & - \textrm{Im} \ln \sn(\theta x + i \psi y, k).\nonumber
\end{eqnarray}
Thus, this new surface can still be thought of as a sum of screw dislocations in the $xy$-plane where we apply a nonlinear transformation to the $z$ coordinate rather than a simple rescaling by a factor $\gamma$.  Notice that $\Phi$ is no longer harmonic because of the functional dependence on $z$.

\section{Topological constructions for chiral twist-grain boundaries}
\label{sec:moire}
\begin{figure}[t]
\begin{center}
\resizebox{3.5in}{!}{\includegraphics{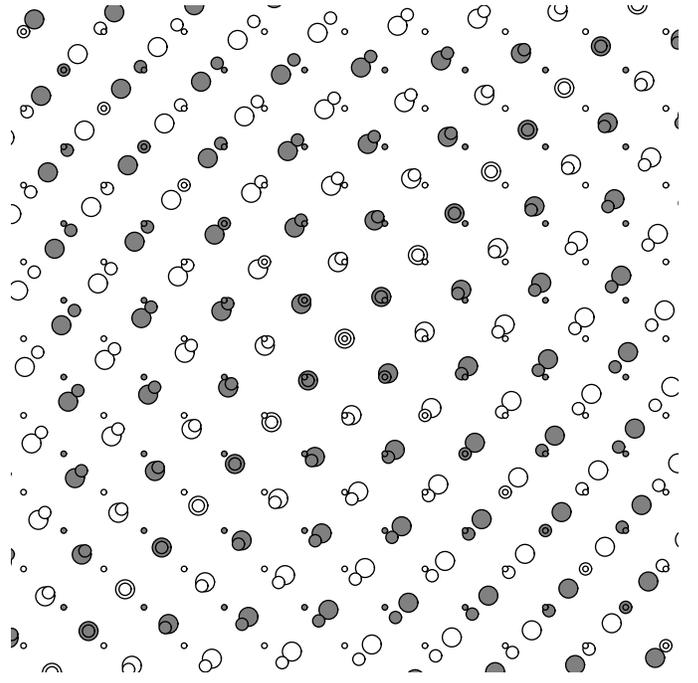}}
\caption{Smectic defects along the pitch axis for a twist angle $\alpha = \pi/2 - \sqrt{2}/32$.  Positive dislocations are open circles and negative dislocations are filled.  The circle radius increasing with each grain boundary.  Two iterations are shown in the figure.}
\label{fig:nonmoire}
\end{center}
\end{figure}
The $\pi/2$ TGB phase structure is not chiral, and our construction of summing dislocations along the $z$ axis (along which the smectic is periodic) necessarily fails to produce any chiral structures.  We can explore TGB phases with twist angles $\alpha < \pi/2$ by applying an additional twist to Schnerk's first surface.  From this point of view, it is more natural to think of the dual defects along the pitch axis, along $\hat{y}$ in our conventions, which are compelled to twist around to follow the surface.  If we make an analogy between the screw dislocations along the pitch axis and a columnar phase of polymers, then the twisting of the surface results in the dislocations braiding around each other in analogy with the polymers in the columnar moir\'e phase~\cite{moire}.

We first briefly review the basic features of the moir\'e phase, which arises when polymer chirality competes with the columnar order.  Much as occurs in the TGB smectic phase, chirality can be incorporated into the polymer lattice by introducing an ordered arrangement of screw dislocations.  In reference~\cite{moire}, the authors consider two cases: either the dislocations form tilt-grain boundaries in which entire rows of columns slide past each other, or the dislocations form a honeycomb lattice which allows groups of polymers to twist around independently.  Here however, the lattice we consider is composed of two kinds of ``polymers'' -- the positive and negative screw dislocations of the
smectic.  To prevent premature aging of the authors and the reader we will refer to the original screw dislocations as ``smectic dislocations'' and the screw dislocations in the columnar lattice of the smectic dislocations as ``columnar dislocations''.  There will not be any point where the distinction will be
clear by context.  
\begin{figure}[b]
\begin{center}
\resizebox{3.5in}{!}{\includegraphics{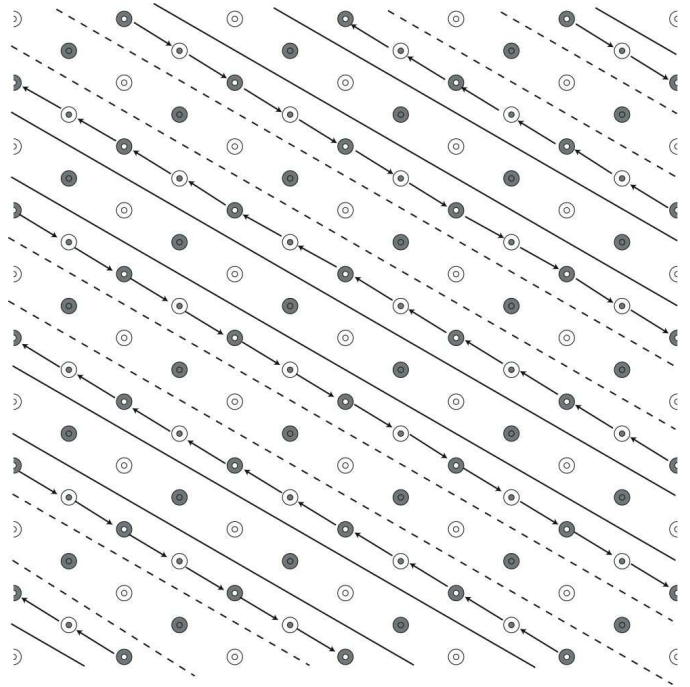}}
\caption{Smectic dislocations along the pitch axis for a twist angle $\alpha = \pi/3$ for tilt-grain boundary structure.  Positive dislocations are open circles and negative dislocations are filled.  The circle radius increases with each grain boundary.  Shown are the dislocations of only two adjacent grain boundaries.  The solid and dashed lines are columnar dislocations in the defect lattice with opposite topological charge.  In the next pair of adjacent grain boundaries, the columnar dislocations rotate by an angle $\pi/3$.  The arrows indicate the relative position from the region below the plane of the page to the region above the plane of the page.}
\label{fig:60moire2}
\end{center}
\end{figure}

\begin{figure}[t]
\begin{center}
\resizebox{3.5in}{!}{\includegraphics{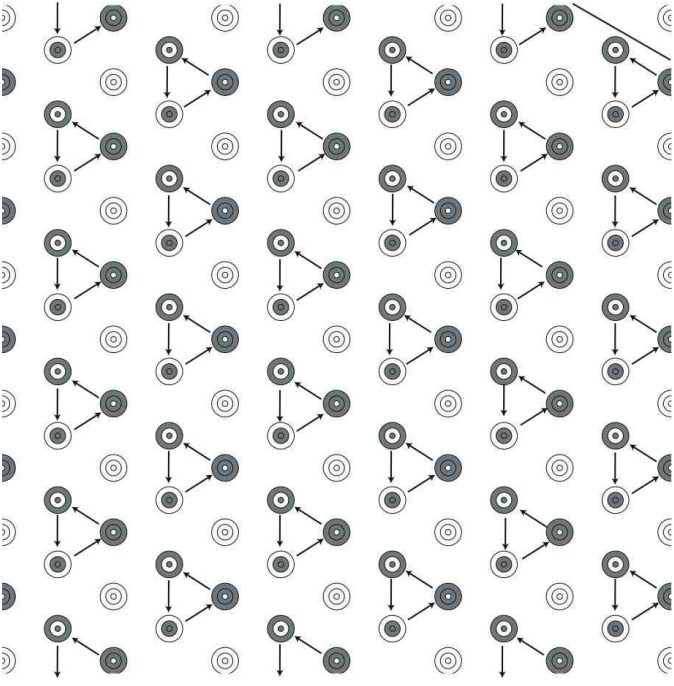}}
\caption{Smectic dislocations along the pitch axis for a twist angle $\alpha = \pi/3$.  Positive dislocations are open circles and negative dislocations are filled.  The circle radius increases with each grain boundary.  }
\label{fig:60moire}
\end{center}
\end{figure}

\begin{figure}[b]
\begin{center}
\resizebox{3.5in}{!}{\includegraphics{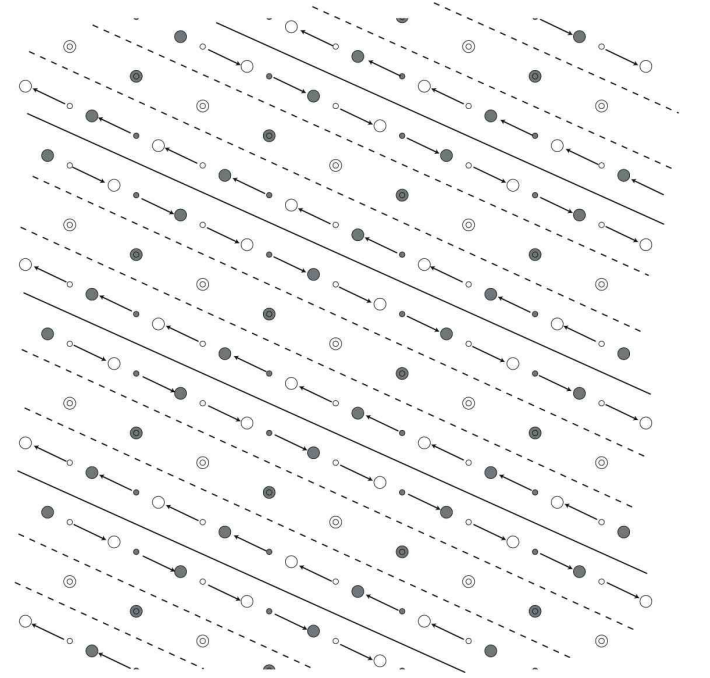}}
\caption{Smectic dislocations along the pitch axis for a twist angle $\alpha = 2 \cot^{-1} \sqrt{5}$ (corresponding to $n=2$).  Positive defects are open circles and negative defects are filled.  The circle radius increases with each grain boundary.  Shown are the defects of only two adjacent grain boundaries.  The solid and dashed lines are columnar dislocations in the defect lattice with opposite topological charge.  In the next pair of adjacent grain boundaries, the columnar screw dislocations rotate by an angle $\alpha$.  The arrows depict the shifts of the columns from below to above the page.  }
\label{fig:48moire2}
\end{center}
\end{figure}

\begin{figure}[t]
\begin{center}
\resizebox{3.5in}{!}{\includegraphics{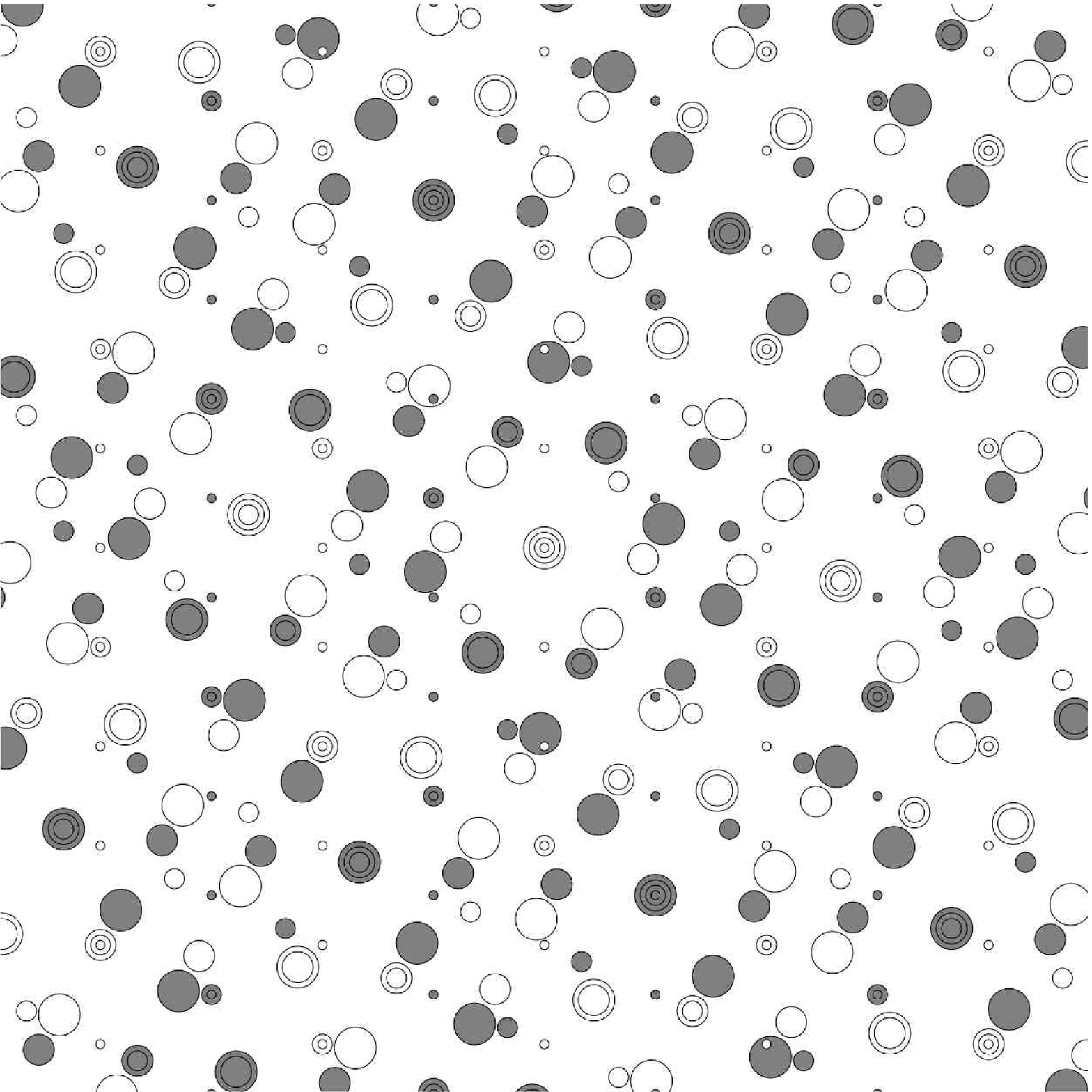}}
\caption{Smectic dislocations along the pitch axis for a twist angle $\alpha = 2 \cot^{-1} \sqrt{5}$ (corresponding to $n=2$).  Positive defects are open circles and negative defects are filled.  The circle radius increases with each grain boundary.  Shown are the smectic dislocations for four adjacent slabs.}
\label{fig:48moire}
\end{center}
\end{figure}

To simplify our analysis to its most basic level, we consider the limit that the grain boundaries are far apart so that we may use the $k=0$ results for the rotation angle of the grain boundaries.   The positive
 smectic defects along the pitch axis sit at
\begin{equation}
\left[z^+_{n m},x^+_{n m}\right] = \left[\ell_z n, \ell_d (m+\hbox{${1\over 2}$}) \right],
\end{equation}
and the negative at
\begin{equation}
\left[z^-_{n m},x^-_{n m} \right]= \left[\ell_z (n+\hbox{${1\over 2}$}),\ell_d m \right ]
\end{equation}
where $\ell_z \equiv b/(2\gamma) = b/[2\cos(\alpha/2)]$, $\ell_d=b/[2\sin(\alpha/2)]$ (as before), and $n,m\in\mathbb{Z}$. For rotation angles $\alpha = \pi/2$, $\ell_z = \ell_d$ and the smectic dislocations form a pair of interwoven square lattices with $+b$ smectic dislocations on one lattice and $-b$ smectic dislocations on a square lattice shifted by half a lattice spacing in each direction. 
To form a twist-grain-boundary phase with rotation angle different from $\pi/2$, we untwist the structure along its pitch axis ({\sl i.e.} the $y$-axis).  In doing this we change the rotation angle $\alpha$.  At the same time, the smectic dislocations that run along the $y$-axis get twisted and the dislocations must bend to accommodate this mismatch.  Though the smectic dislocations interact, acquire edge components, and suffer any number of other deformations, we still expect that when the rotation from one slice of the lattice to the next allows some fraction of the smectic dislocations to remain straight that there will be a local minimum in the free energy.  These preferred rotations will generate a structure analogous to the moir\'e phases of chiral polymers.  In the case of $\alpha=\pi/2$, we have a perfect agreement of the lattices as the smectic dislocations are straight lines along $\hat y$.  For small rotations away from $\pi/2$, as shown in Fig. \ref{fig:nonmoire}, the energy grows.  Thus for a highly chiral mesogen which favors grain boundaries with rotation angles close to $\pi/2$, this local lock-in mechanism could very well force the formation of the non-chiral $\pi/2$ grain boundary.  Note, however, that the nematic director can and will continue to rotate with a preferred handedness -- it is only the layer structure
that is achiral.  

At other angles, we must introduce screw dislocations into the columnar lattice of smectic dislocations.    These columnar dislocations will effect the necessary rotations of the lattice.  In general, if we have a rectangular lattice with lattice constants $\ell_z$ and $\ell_d$, there will be a lock-in angle when we rotate so that the lattice point at $\left[(n+\hbox{$1\over 2$})\ell_z,\ell_d/2\right]$ rotates onto $\left[(n+\hbox{$1\over 2$})\ell_z,-\ell_d/2\right]$, and
\begin{equation}
\tan(\beta_n/2) =\frac{ \hbox{${1\over 2}$}\ell_d}{(n+\hbox{$1\over 2$})\ell_z}
\end{equation}
Note, however, that $\ell_d/\ell_z=\cot(\alpha/2)$ depends on the angle of rotation.  Consistency requires that $\alpha=\beta_n$ and we 
find
\begin{equation}\label{angles}
\tan(\beta_n/2) = \frac{1}{\sqrt{2n+1}},
\end{equation}
from which we get the first few moir\'e angles, $\beta_0=\pi/2$, $\beta_1=\pi/3$, and $\beta_2=\tan^{-1}(1/\sqrt{5})\approx 48.19^\circ$.   Remarkably, experiments on large angle grain boundary phases \cite{clark,clarkprivate} have observed $90^\circ$, $60^\circ$ and $45^\circ$ grain
boundaries, close to, if not exactly, the sequence we find here.  Note also that the precise value of the angles is altered by adjacent grain boundaries; the rotation angle we found in (\ref{twist}) is
\begin{equation}
\sin\left(\frac{\alpha}{2}\right) = \left[\frac{b\sqrt{1-k^2}\K(k)}{(\pi\ell_d)}\right]\approx\frac{b}{2\ell_d}\left[1-4q  +\ldots\right]
\end{equation} 
so that $\ell_d\approx (b/[2\sin(\alpha/2)])(1-4q)$.  This modifies (\ref{angles}) so that
\begin{equation}
\tan{\beta_n/2} \approx \frac{1+4e^{-2\pi\ell_b/\ell_d}}{\sqrt{2n+1}}
\end{equation}
for our structures.  The effect is both in the wrong direction ({\sl i.e.}, making the predicted angles larger) and extremely small
($e^{-2\pi}\approx 10^{-3}$), so pure geometry is not likely to explain the observed twist angles and the (small) energetic effects must
be included.  
 
For $n=1$, $\theta_n = \pi/3$, which is especially simple because the interlocking square lattices becomes a single triangular lattice: $\ell_z/\ell_d = \sqrt{3}$.   Though the lattice is invariant
under rotations by $\pi/3$, this is only the case when we ignore the sign of the smectic dislocations.  The handedness of the
rotation and the necessity of columnar dislocations arises because of the alternating signs.  In Fig. \ref{fig:60moire2}, we  
show the arrangement of columnar dislocations necessary to rotate the smectic dislocations.  Adjacent {\sl pairs} of columnar dislocations have the opposite topological charge, so the net charge remains zero from the point of view of the undecorated lattice.   This arrangement slides every other line of smectic screw dislocations  relative to each other in opposite directions.  Though it is not chiral, chirality appears because, in adjacent columnar grain boundaries, the columnar dislocations must rotate along with the smectic layers and dislocations.  After three of these rotations the smectic dislocations return to their original arrangement.   In Fig. \ref{fig:60moire} we
show the unit cell of the moir\'e phase of the smectic dislocations.

For $n=2$ ($\alpha = 2 \cot^{-1} \sqrt{5} \approx 48.19^\circ$) we consider Figs. \ref{fig:48moire2} and \ref{fig:48moire}.  As with the usual moir\'e phase of
columnar liquid crystals, there is a finite fraction of smectic dislocations below the grain boundary which do not match up precisely with a smectic dislocation above the grain boundary, even ignoring the sign of the smectic dislocation.  In Fig. \ref{fig:48moire2} we depict
a single grain boundary.   As before, the columnar dislocations in the next slab will rotate by  $2\cot^{-1}\sqrt{5}$.  Though there is a coincident subset of lattice points in each pair of adjacent grain boundaries, this set changes as we move along the pitch axis.  In Fig. \ref{fig:48moire} we draw one, very complicated, piece of the structure composed of just four columnar slabs.  As with the traditional moir\'e phases, we believe that the allowed
rotation angles are all irrational fractions of $2\pi$, but for the cases of $n=0$ or $n=1$.  This lack of a repeating pattern persists for the higher moir\'e angles giving moir\'e/twist-grain-boundary structures of increasing complexity.  The mind boggles.

In closing, we conclude that the $\pi/3$ structure is particularly simple from the point of view of the $\pi/2$ structure.  Having twisted Schnerk's surface to give $\pi/3$ grain boundaries, continuing to change the angle raises the energy again from these purely geometric considerations.  It is interesting to note that some materials in which $\pi/2$ TGB phases have been observed also contain regions of $\pi/3$ TGB phases~\cite{clark}.  Our results suggest that the prevalence of these two rotation angles is due to the apparent simplicity of the defect network along the TGB pitch axis  -- other angles presumably raise the energy of the structure.

\section{Discussion}
\label{sec:discussion}
In summary, we have presented an explicit analytical construction for the $\pi/2$ twist-grain-boundary structure by directly summing screw dislocations.  This yields a phase field from which the energetics of the structure can be estimated.  Our construction relies crucially on a duality: dislocations within one grain boundary can be rotated by ninety degrees, as long as the topological charge is simultaneously reversed, without changing the layers.  For a twist angle of precisely $\pi/2$, the dislocations can be made into a parallel, bipartite lattice of positive and negative dislocations.

From this construction emerges another unexpected symmetry; we may view the TGB structure as composed of a bipartite lattice of defects {\sl along} the pitch axis.  For the $\pi/2$ structure, these defects are straight.  Applying an additional twist to the $\pi/2$ structure yields a twist-grain-boundary phase with twist angle $\alpha < \pi/2$.  The defects along the pitch axis must then twist along with the structure, and geometric considerations suggest that this raises the energy of the structure.  However, we have identified a set of twist angles $\alpha$ for which the screw dislocations can be made partially straight, in analogy to the moir\'e phase of columnar liquid crystals~\cite{moire}.  The structure for $\alpha=\pi/3$ is particularly simple, consisting of triplets of defects braiding around each other.

We conclude by noting that Schnerk's first surface is an extremum of the linearized smectic free energy.  Though we evaluated its energy with the nonlinear elastic strain, it is not clear how nonlinearities will modify the structure of the layers and the interaction between defects.  In the case of edge dislocations, the defect interactions are not strongly modified by the presence of nonlinearities (though the layers are) \cite{bps}, suggesting that our conclusions are robust.  Finally, we point out that large twist angles are likely to result in a decoupling of the director from the layer normal.   It is also tempting to attempt to identify edge dislocations in the smectic with screw dislocations in the columnar crystal of defects, and {\sl vice versa}.    There is much to do.

\begin{acknowledgments}
We thank M. Cohen, G. Gibbons, R. Kusner, and T.C. Lubensky for interesting discussions. This work was supported through NSF Grant DMR05-47230, the Donors of the ACS Petroleum Research Fund, and a gift from L.J. Bernstein.
\end{acknowledgments}

\appendix*
\section{Identities for $k^2<0$}
Here we establish some needed identities for the case of $k$ pure imaginary, {\sl i.e.} $k^2\in\mathbb{R}$ and $k^2<0$.  Though these are surely known, we put them here for completeness.  

The first result we need is $\hbox{Im}\,\K'(k) = - \K(k)$.  We start with the definition of $\K'(k)$:
\begin{equation}
i\K'(k) \equiv \int_1^{1/ k} {dx \over\sqrt{\left(1-x^2\right)\left(1-k^2x^2\right)}}
\end{equation}
Setting $k=i\kappa$, for $\kappa\in\mathbb{R}^+$, we have
\begin{equation}
i\K'(i\kappa) = \int_1^{-i/ \kappa} {dz\over\sqrt{\left(1-z^2\right)\left(1+\kappa^2z^2\right)}}
\end{equation}
We choose three branch cuts in the complex plane as follows: the first connects $-1$ to $1$ along the
real axis.  The second starts at $i/\kappa$ and runs up the imaginary axis, while the third starts
at $-i/\kappa$ and runs down the imaginary axis.  
We choose a contour from $1$ to $-i/\kappa$ which goes from $1$ to $0$ along the real axis 
and then goes from $0$ to $-i/\kappa$ along the imaginary axis.  We have
\begin{eqnarray}
i\K'(i\kappa)&=&-\int_1^0 {dx\over\sqrt{\left(1-x^2\right)\left(1+\kappa^2x^2\right)}}\nonumber\\&& 
-i\int_0^{-1/\kappa} { dy\over\sqrt{\left(1+y^2\right)\left(1-\kappa^2y^2\right)}}
\end{eqnarray}
where the overall minus comes from being underneath the first cut.
Since both integrands are real, it follows that $\hbox{Im}\,\K'(i\kappa) = -\int_0^1 dx\left[(1-x^2)(1+\kappa^2x^2)\right]^{-1/2}\equiv -\K(i\kappa)$.

Next we show that, again,  for $k^2\in\mathbb{R}^-$, $\vert \cs(u,k)\,\dn(u,k) \vert^2 = 1-k^2$ along $u=\K(k)/2 + it$ and $u=t + i\hbox{Re}\,\K'(k)/2$.  We use the following expressions for $\sn(u)$, $\cn(u)$ and $\dn(u)$ in terms of the Jacobi Theta functions (using the conventions in \cite{bateman})
\begin{eqnarray}
\sn(u,k) &=& \frac{\theta_3(0\vert\tau)\theta_1(v\vert \tau)}{\theta_2(0\vert\tau)\theta_4(v\vert \tau)}\nonumber\\
\cn(u,k) &=& \frac{\theta_4(0\vert\tau)\theta_2(v\vert \tau)}{\theta_2(0\vert\tau)\theta_4(v\vert \tau)}\nonumber\\
\dn(u,k) &=& \frac{\theta_4(0\vert\tau)\theta_3(v\vert \tau)}{\theta_3(0\vert\tau)\theta_4(v\vert \tau)}
\end{eqnarray}
where $v=u/[2\K(k)]$. 
from which it follows that 
\begin{eqnarray}
I(u,k)&\equiv&\cs(u,k)\,\dn(u,k) \\ \nonumber
&=&\frac{\cn(u,k)\,\dn(u,k)}{\sn(u,k)}\\ \nonumber
&=&\frac{\theta_4^2(0\vert\tau)}{\theta_3^2(0\vert\tau)}\frac{\theta_2(v\vert \tau)\theta_3(v\vert \tau)}
{\theta_4(v\vert \tau)\theta_1(v\vert \tau)}
\end{eqnarray}
 We have $\sqrt{1-k^2} = \theta_4^2(0\vert\tau)/\theta_3^2(0\vert\tau)$ \cite{bateman} so that when $k^2$ is real and negative we find
\begin{eqnarray}
\vert I(u,k)\vert^2 &=& (1-k^2)\left\vert\frac{\theta_2(v\vert \tau)\theta_3(v\vert \tau)}
{\theta_4(v\vert \tau)\theta_1(v\vert \tau)}
\right\vert^2
\end{eqnarray}
Rewriting all the Theta functions in terms of $\theta_3(v\vert\tau)$, we have
\begin{eqnarray}
\frac{\vert I(u,k)\vert^2}{1-k^2} &=& \left\vert\frac{\theta_3(v\vert \tau)\theta_3(v+{1\over 2}\tau\vert \tau)}
{\theta_3(v+{1\over 2}\vert \tau)\theta_3(v+{1\over 2}+{1\over 2}\tau\vert \tau)}\right\vert^2\nonumber\\
\end{eqnarray}
Fortuitously, $\theta_3(v\vert\tau)$ can be 
expressed in terms of the Jacobi triple product:
\begin{widetext}
\begin{eqnarray}
\theta_3(v\vert\tau) = \prod_{m=1}^\infty \left[1-e^{2\pi i m\tau}\right]\left[1+e^{(2m-1)\pi i\tau + 2\pi iv}\right]\left[1+e^{(2m-1)\pi i\tau - 2\pi iv}\right]
\end{eqnarray}\end{widetext}
Setting $u=\K(k)/2 + it$ ($t\in\mathbb{R}$), makes $v={1\over 4} + i t/[2\K(k)]$. Since $\K(k)$ is real, the real part of $v$ is set.  Moreover,  $\tau = 1+i\frac{{\rm Re}\K'(k)}{\K(k)}$ also has a prescribed real part.   Letting $p=e^{-\pi \hbox{Re}\K'(k)/\K(k)}$ and $g=e^{-\pi t/\K(k)}$ we have:
\begin{widetext}
\begin{equation}\label{eq:id1}
\frac{\vert I(\K(k)/2 + it,k)\vert^2}{1-k^2} = \left\vert
\frac{\prod_{m=1}^\infty\left[1 -ip^{2m-1}g\right]\left[1+ip^{2m-1}g^{-1}\right]\left[1+ip^{2m}g\right]\left[1-ip^{2m}g^{-1}\right] }{\prod_{m=1}^\infty\left[1+ip^{2m-1}g\right]\left[1-ip^{2m-1}g^{-1}\right]\left[1-ip^{2m}g\right]\left[1+ip^{2m}g^{-1}\right]}\right\vert^2=1
\end{equation}
\end{widetext}
since $\vert a/b\vert = \vert a/b^*\vert$.

Similarly, when $u=t+i\hbox{Re}\K'(k)/2$, $v= t/[2\K(k)] +(\tau -1)/4$.  Setting $h=e^{i\pi t/\K(k)}$ we have
\begin{widetext}
\begin{eqnarray}\label{eq:id2}
\frac{\vert I(t+ i\hbox{Re}\K'(k)/2,k)\vert^2}{1-k^2} &=& \left\vert
\frac{\prod_{m=1}^\infty\left[1 -p^{2m-{1\over 2}}h\right]\left[1-p^{2m-{3\over 2}}h^*\right]\left[1+p^{2m+{1\over 2}}h\right]\left[1+p^{2m-{5\over 2}}h^*\right] }{\prod_{m=1}^\infty\left[1+p^{2m-{1\over 2}}h\right]\left[1+p^{2m-{3\over 2}}h^*\right]\left[1-p^{2m+{1\over 2}}h\right]\left[1-p^{2m-{5\over 2}}h^*\right]}\right\vert^2\nonumber\\
&=&  \left\vert
\frac{\prod_{m=1}^\infty\left[1 -p^{2m-{1\over 2}}h\right]\left[1-p^{2m-{3\over 2}}h^*\right]\left[1+p^{2m+{1\over 2}}h\right]\left[1+p^{2m-{5\over 2}}h^*\right] }{\prod_{m=1}^\infty\left[1-p^{2m-{5\over 2}}h\right]\left[1-p^{2m+{1\over 2}}h^*\right]\left[1+p^{2m-{3\over 2}}h\right]\left[1+p^{2m-{1\over 2}}h^*\right]}\right\vert^2\nonumber\\
&=&\left\vert\frac{\left(1-p^{{1\over 2}}h^*\right)\left(1+p^{-{1\over 2}}h^*\right)}{\left(1-p^{-{1\over 2}}h\right)\left(1+p^{1\over 2}h\right)}\right\vert^2=1
\end{eqnarray}
\end{widetext}
where the final equality follows from multiplying by $\vert\frac{p^{1/2}h}{p^{1/2}h^*}\vert^2=1$.  Writing
$\zeta = \theta x + i\psi y=2\K(k)x/\ell_d + i\hbox{Re}\K'(k)y/\ell_b$, identities (\ref{eq:id1}) and (\ref{eq:id2}) show that, when $\theta=\psi$, the compression
$\left(\nabla\Phi\right)^2$ is constant along the lines $x=\ell_d/4$ and $y=\ell_b/2$, plus all their periodic
translates.  These are precisely the lines that run {\sl between} the rows of screw dislocations.


\begin{thebibliography}{99}
\bibitem{Kafka} But Kafka admonishes us that ``The right understanding of any matter and a misunderstanding of the same matter do not wholly exclude each other.''  From {\sl The Trial}.
\bibitem{footnote1} Another symmetric construction is found by replacing the elliptic functions with standard trigonometric functions.  This gives the commonly used nodal approximation for the Schwarz D surface, which also arises in an appropriate $k \rightarrow 0$ limit of the elliptic form.
\end{thebibliography}
\end{document}